\documentclass[prd,10pt,aps,twocolumn,superscriptaddress,showpacs,nofootinbib,floatfix,amssymb,amsmath]{revtex4-1}
\usepackage{rotating}
\usepackage{multirow}

\newcommand{\DD}{\mathcal{D}}

\newcommand{\al}[1]{\begin{align}#1\end{align}}

\newcommand{\braket}[1]{\Bigl\langle #1 \Bigr \rangle}
\newcommand{\sbraket}[1]{\bigl\langle #1 \bigr\rangle}

\newcommand{\xx}{x}
\newcommand{\yy}{y}
\newcommand{\kk}{k}

 \newcommand{\beqn}{\begin{eqnarray}}
 \newcommand{\eeqn}{\end{eqnarray}}
\newcommand{\eq}[1]{(\ref{#1})}

\newcommand{\cO}{{\cal O}}

\newcommand{\bs}{\boldsymbol}
\newcommand{\Z}{{\mathbb Z}}

\newcommand{\beqs}{\begin{subequations}}
\newcommand{\eeqs}{\end{subequations}}

\begin{document}

\title{Phonon spectrum of QCD vacuum in magnetic-field-induced superconducting phase}

\author{M. N. Chernodub}\thanks{On leave from ITEP, Moscow, Russia.}
\affiliation{CNRS, Laboratoire de Math\'ematiques et Physique Th\'eorique, Universit\'e Fran\c{c}ois-Rabelais Tours,\\ F\'ed\'eration Denis Poisson, Parc de Grandmont, 37200 Tours, France}
\affiliation{Department of Physics and Astronomy, University of Gent, Krijgslaan 281, S9, B-9000 Gent, Belgium}
\author{Jos Van Doorsselaere}
\affiliation{CNRS, Laboratoire de Math\'ematiques et Physique Th\'eorique, Universit\'e Fran\c{c}ois-Rabelais Tours,\\ F\'ed\'eration Denis Poisson, Parc de Grandmont, 37200 Tours, France}
\affiliation{Department of Physics and Astronomy, University of Gent, Krijgslaan 281, S9, B-9000 Gent, Belgium}
\author{Henri Verschelde}
\affiliation{Department of Physics and Astronomy, University of Gent, Krijgslaan 281, S9, B-9000 Gent, Belgium}

\begin{abstract}
In the background of a sufficiently strong magnetic field the vacuum was suggested to become an ideal electric conductor (highly anisotropic superconductor) due to an interplay between the strong and electromagnetic forces. The superconducting ground state resembles an Abrikosov lattice state in an ordinary type--II superconductor: it is an inhomogeneous structure made of a (charged vector) quark-antiquark condensate pierced by vortices. In this paper the acoustic (phonon) vibrational modes of the vortex lattice are studied at zero temperature. Using an effective model based on a vector meson dominance, we show that in the infrared limit the longitudinal (transverse) acoustic vibrations of the vortex lattice possess a linear (quadratic) dispersion relation corresponding to  type I (type II) Nambu--Goldstone modes. 
\end{abstract}

%Quantum chromodynamics, 12.38.-t
%Electromagnetic interactions, 13.40.-f
%Superconductivity, new topics in, 74.90.+n

\date{\today}

\maketitle

\section{Introduction}

The quantum vacuum may exhibit quite unusual properties in a strong magnetic field background if the magnetic field exceeds the hadronic-scale, $e B \gtrsim \Lambda_{\mathrm{QCD}}^2$. A strong magnetic field enhances the chiral symmetry breaking in QCD due to the magnetic catalysis phenomena~\cite{ref:magnetic:catalysis}. As a result, magnetic field affects the finite-temperature phase diagram of  QCD~\cite{ref:Tc:rising} by shifting the transition temperature in a quite unexpected way~\cite{ref:lattice:results}. The phase structure of the QCD quark matter is also very sensitive to the presence of a strong magnetic field~\cite{ref:magnetic:matter}. In addition, the magnetized quark matter should exhibit new transport phenomena, such as the chiral magnetic effect~\cite{ref:CME,ref:CME:cond:matt}, while the magnetic field background may affect standard transport phenomena~\cite{ref:transport:related}.

The interest in physics of extreme magnetic fields is justified by the fact that such strong fields may emerge in noncentral heavy-ion collisions. For example, in lead-lead collisions at the LHC, the strength of the generated magnetic field may reach  $B \sim 10^{16} \,{\mathrm{T}}\sim  70 m_\pi^2/e$~\cite{ref:field:estimation}. The magnetic field may affect the quark-gluon plasma created by overlapping heavy ions as well as the vacuum between the ions if these ions near-miss each other in ultraperipheral collisions.

It was recently suggested that the strong magnetic field may cause the QCD vacuum to behave as an anisotropic perfect conductor (superconductor) at low temperatures~\cite{Chernodub:2010qx,Chernodub:2011mc} if the strength of the magnetic field exceeds the critical value
\beqn
B_c \simeq 10^{16} \, {\mathrm{Tesla}}
\qquad  {\mathrm{or}} \qquad
e B_c \simeq 0.6\,\mbox{GeV}^2\,.
\quad
\label{eq:Bc}
\eeqn 
The transition from the insulating to superconducting regimes is caused by the condensation of the quark-antiquark pairs, which carry the quantum numbers of the electrically charged $\rho$ mesons (we call it later ``$\rho$-meson condensate''). The condensate is an anisotropic and inhomogeneous structure which may lack, in thermodynamics sense, a local order parameter beyond the mean field approximation~\cite{ref:comment}. 

In contrast to the standard phenomenon of superconductivity the magnetic field penetrates the $\rho$ condensate: the  Meissner effect is absent in the new phase because of the vector rather than scalar nature of the condensate~\cite{Chernodub:2010qx}. 

{This phenomenon has a known counterpart in the solid state physics which is sometimes the ``reentrant'' superconductivity~\cite{ref:Zlatko}. In an increasing magnetic field a type-II superconductor will eventually experience a transition to a normal phase and may then -- according to the proposal of Ref.~\cite{ref:Zlatko} --  ``reenter'' the superconducting regime again. The reentrant superconductivity is associated with an inhomogeneous $(p+ip)$ condensation of electron pairs and is suggested to be realized in certain materials. In the reentrant regime the emergent superconductivity does not screen the background magnetic field, while the breaking of the electromagnetic $U(1)$ symmetry results in appearance of short-range modulations (inhomogeneities) of the condensate in the transverse plane.  In scope of the $\rho$--meson condensation the absence of the Meissner effect was discussed in Refs.~\cite{Chernodub:2010qx}, and a related proposal for ferromagnetic superconductors is put forward in Ref.~\cite{ref:ferromagnetic}.}

Generally, a possible appearance of the $\rho$-meson condensate in strong magnetic field is not very surprising: it is quite similar to the gluon condensation in QCD~\cite{ref:YM} and to the $W$-boson condensation in the electroweak model~\cite{ref:EW}. The question on the possibility of $\rho$--meson condensation in high magnetic fields in QCD was also briefly raised in Ref.~\cite{ref:earlier}. Both $\rho$ mesons, gluons and $W$ bosons are vector particles, which are sensitive to high magnetic (chromomagnetic, in the case of gluons) fields. The closely related question of condensation of charged vector particles in the background of a magnetic field is discussed in Ref.~\cite{ref:related:vector}.

{The condensation of quark-antiquark pairs with $\rho$--meson quantum numbers were also found in various holographic approaches~\cite{ref:holographic,ref:holographic:hexagonal}, in local~\cite{Chernodub:2011mc} and nonlocal~\cite{Frasca:2013kka} Nambu--Jona-Lasinio (NJL) models} as well as in numerical simulations of quenched lattice QCD~\cite{Braguta:2011hq}. Due to the anisotropic superconductivity the vacuum may become an exotic hyperbolic metamaterial which shares a similarity with diffractionless ``perfect lenses''~\cite{Smolyaninov:2011wc}. The $\rho$--meson condensation is a subject of an ongoing discussion~\cite{ref:comment,ref:discussion}.

In the mean field approach the vortices form a hexagonal lattice which is similar to the mixed Abrikosov state in an ordinary type--II superconductor. Since the vortex lattice breaks the continuous spatial symmetries, it should give rise to Nambu-Goldstone modes. These massless modes are, in fact, phonons which correspond to elementary vibrational excitations of the vortex lattice. The presence of phonons is crucial for understanding the stability of the vortex lattice against quantum and thermal fluctuations: {depending on the spectrum of the phonon modes} the vortex lattice may experience a chain of deformations and, eventually, melt into a vortex liquid~\cite{ref:type-II:Review}. 

{According to the numerical simulations of lattice QCD in strong magnetic field, the vortices appear in a liquid phase rather then in a form of the ordered vortex crystal~\cite{Braguta:2013uc}. Since the phonon modes are responsible for the melting of the vortex lattice (crystal), the study of the vortex excitations is physically interesting.} In this paper we describe the vibrations of the $\rho$-vortex lattice following a well-developed calculation method for the phonon spectrum in Abrikosov vortex lattices in type II superconductors. A good review on this subject can be found in Ref.~\cite{ref:type-II:Review}.  

{In order to address the problem of the phonon spectrum of the vortex lattice we work in the framework of an effective electrodynamics for the $\rho$ mesons~\cite{Djukanovic:2005ag} based on the vector meson dominance~\cite{Sakurai:1960ju}. This model describes the superconducting ground state in consistency with other effective approaches~\cite{ref:holographic,ref:holographic:hexagonal,Chernodub:2011mc,Frasca:2013kka}. The choice of this model is also justified by the analogy with usual superconductivity, where the standard Ginzburg-Landau approach describes well both the vortex-lattice ground state and its phonon excitations. Finally, we would like to mention that the use of the effective field theories in highly-magnetized zero-temperature QCD is generally supported by numerical lattice calculations. The relevant examples include a linear behavior of the chiral condensate (Refs.~\cite{Shushpanov:1997sf} and \cite{Bali:2012zg}, respectively) and a peculiar nonlinear behavior of the chiral magnetization (\cite{Frasca:2011zn} and \cite{Buividovich:2009ih} respectively) as functions of magnetic field.}

{The structure of the paper is as follows. The effective electrodynamics of $\rho$ mesons and the ground state} of this model in strong magnetic field are described in Section~\ref{sec:model}. We introduce the basis of crystal wavefunctions and study their basic properties in Section~\ref{sec:basis}. In Section~\ref{sec:phonons} we derive the dispersion relation for the phonons. We show that in the infrared limit the transverse acoustic vibrations of the $\rho$-vortex lattice possess a quadratic ``super-soft' dispersion relation corresponding to  type II Goldstone modes, while the longitudinal modes are always linear in momentum similarly to type I Goldstone bosons. Our main result for the dispersion relation for the low-energy phonons is given in Eq.~\eq{eq:omega:k:low}:
\beqn
\omega_k^2 & = & k_z^2 + f(B) \left({\bs k}^2\right)^2 + \dots\,,
\label{eq:result}
\eeqn
where $k_z$ and $\bs k$ are the longitudinal and transverse momenta, respectively, and the field-dependent prefactor $f(B)$ is given explicitly in Eq.~\eq{eq:f:B}. Apart from the prefactor, this dispersion relation for the low-energy phonons in the $\rho$-vortex ground state has the same qualitative form as the dispersion relation for the acoustic phonons in Abrikosov vortex lattices in conventional superconductors. {The presence of the supersoft (quadratic) transversal mode in the phonon spectrum~\eq{eq:result} may be responsible for the melting of the mean-field vortex lattice into the vortex liquid. The latter state was indeed observed in lattice simulations~\cite{Braguta:2013uc}.} Our conclusions are given in the last section.

\section{Model, phases and approximations}
\label{sec:model}

The $\rho$ mesons are charged and neutral vector particles made of light ($u$ or $d$) quarks and antiquarks. A self-consistent quantum electrodynamics for the $\rho$ mesons can be described by the Djukanovic--Schindler--Gegelia--Scherer (DSGS) Lagrangian~\cite{Djukanovic:2005ag}:
\beqn
{\cal L} & = &-\frac{1}{4} \ F_{\mu\nu}F^{\mu\nu}
- \frac{1}{2} \ \rho^\dagger_{\mu\nu}\rho^{\mu\nu} + m_\rho^2 \ \rho_\mu^\dagger \rho^{\mu}
\label{eq:L:rho}\\
&& -\frac{1}{4} \ \rho^{(0)}_{\mu\nu} \rho^{(0) \mu\nu}+\frac{m_\rho^2}{2} \ \rho_\mu^{(0)}
\rho^{(0) \mu} +\frac{e}{2 g_s} \ F^{\mu\nu} \rho^{(0)}_{\mu\nu}\,,
\nonumber
\eeqn
where 
\beqn
\rho_\mu \equiv \rho^- = \frac{\rho^{(1)}_\mu - i \rho^{(2)}_\mu}{\sqrt{2}}\,,
\eeqn
and $\rho^+_\mu = \rho^\dagger_\mu$ are charged $\rho$ meson fields,  $\rho^{(0)}_\mu$ is the neutral $\rho$ meson field and $A_\mu$ is the electromagnetic field. The field strengths in Eq.~\eq{eq:L:rho} are as follows:
\beqn
F_{\mu\nu} & = & \partial_\mu A_\nu-\partial_\nu A_\mu\,,
\label{eq:F}\\
{f}^{(0)}_{\mu\nu} & = & \partial_\mu \rho^{(0)}_\nu-\partial_\nu \rho^{(0)}_\mu\,,
\label{eq:f0}\\
\rho^{(0)}_{\mu\nu}& = & {f}^{(0)}_{\mu\nu}
- i g_s(\rho^\dagger _\mu \rho_\nu-\rho_\mu\rho^\dagger _\nu)\,,
\label{eq:rho0}\\
\rho_{\mu\nu} & = & D_\mu \rho_\nu - D_\nu \rho_\mu\,,
\label{eq:rho}
\eeqn
where $D_\mu = \partial_\mu + i g_s \rho^{(0)}_\mu - i e A_\mu$ is the covariant derivative and the phenomenological $\rho \pi \pi$ coupling is
\beqn
g_s \equiv g_{\rho\pi\pi} \approx 5.88\,.
\label{eq:gs}
\eeqn
The $\rho$-meson mass at the vanishing magnetic field is $m_\rho (B=0) = 775.5\,\mbox{MeV}$ and the mass of the neutral $\rho^{(0)}$ meson is as follows:
\beqn
m_0 \equiv m_{\rho^{(0)}} = m_\rho \Bigl(1 - \frac{e^2}{g_s^2}\Bigr)^{-\frac{1}{2}}\,.
\label{eq:m:rho0}
\eeqn

The DSGS model \eq{eq:L:rho} is basically the vector meson dominance model~\cite{Sakurai:1960ju} coupled to electromagnetism with the following Abelian gauge transformations:
\beqn
U(1)_{\mathrm{e.m.}}: \quad
\left\{
\begin{array}{lcl}
\rho_\mu(x) & \to & e^{i e \omega(x)} \rho_\mu(x)\,,\\
\rho^{(0)}_\mu(x) & \to & \rho^{(0)}_\mu(x)\,,\\
A_\mu(x) & \to & A_\mu(x) + \partial_\mu \omega(x)\,. \quad
\end{array}
\right.
\label{eq:gauge:invariance}
\eeqn

We consider the model~\eq{eq:L:rho} in a background of a uniform static magnetic field parallel to the $x_3$ axis. The corresponding gauge potential is as follows:
\beqn
\label{eq:Aext}
A_x = - \frac{B }{2} y\,, \quad  A_y = \frac{B }{2} x\,, \quad  A_z = A_t = 0\,. \quad 
\eeqn
We ignore quantum fluctuations of the gauge and meson fields, thus treating the effective model~\eq{eq:L:rho} at the classical level. We always assume $ e B  >0$ for simplicity.

The model~\eq{eq:L:rho} predicts that if the magnetic field exceeds the critical value~\eq{eq:Bc} then the vacuum enters a new phase where positively and negatively charged $\rho$--meson fields condense~\cite{Chernodub:2010qx}. In this phase the vacuum becomes a perfect electric conductor: the electric current can be carried by the $\rho$ condensates along the lines of the magnetic field without dissipation. The transport of electric charge along the magnetic field is described by a one-dimensional London equation, therefore we refer to the perfectly ideally conducting phase as ``the superconductor phase''~\cite{Chernodub:2010qx,Chernodub:2011mc}.

The electric superconductivity of the vacuum in strong magnetic field is caused by a $(p + i p)$--wave condensation of the charged $\rho$--meson field $\rho_\mu(x) \equiv \langle {\bar u}(x) \gamma_\mu d (x)\rangle$ with $\rho = \rho_x = - i \rho_y \neq 0$. Other vector components of the vector condensate are zero, $\rho_z = \rho_t = 0$. In the mean--field approach and at the classical level of the effective model~\eq{eq:L:rho}, the superconducting ground state of the vacuum resembles the Abrikosov lattice state in an ordinary type--II superconductor~\cite{Chernodub:2011mc,Chernodub:2010qx}. 

The mean-field solution for the ground state does not depend on the longitudinal $z$ coordinate. Therefore it is convenient to combine the transverse vectors into complex scalars: $\partial = \partial_x + i \partial_y$,  ${\bar \partial} = \partial_x - i \partial_y$, or in general $\cO = \cO_x + i \cO_y$,  ${\bar \cO} = \cO_x - i \cO_y$. As an exception to this general convention we put
\beqn
\rho=\rho_x-i\rho_y,
\eeqn
as the other combination simply vanishes in the LLL approximation:
\beqn
\rho_x+i\rho_y\equiv 0\qquad ({\mathrm{LLL}}).
\label{eq:rhocomp}
\eeqn

The static (potential) energy density corresponding to the DSGS Lagrangian~\eq{eq:L:rho} is -- assuming (\ref{eq:rhocomp}) -- as follows:
\beqn
& & \mathcal{E}_\perp [\rho,\rho_0,A] = \frac{1}{2}\vert\DD \rho\vert^2+\frac{1}{2}F_{xy}^2+\frac{1}{2}\bigl( \rho^{(0)}_{xy} \bigr)^2\nonumber \\
& & \hskip 9mm + \frac{1}{2}m_\rho^2 \bigl(\vert \rho^{(0)}\vert^2  + \vert\rho\vert^2 \bigr) - \frac{e}{g_s}F_{xy}\rho^{(0)}_{xy}\,,
\label{dsgspot}
\eeqn
where
\beqn
\DD \rho & = & \partial\rho+ig_s\rho^{(0)}{ \rho} - ieA { \rho},\\
F_{xy} &=& -\frac{ i}{2}({\bar \partial} A - \partial {\bar A})\quad = B\ ,\\
\rho^{(0)}_{xy} & = & \frac{g_s}{2} \vert\rho\vert^2-\frac{i}{2} \bigl(\bar\partial\rho^{(0)}-\partial\bar\rho^{(0)}\bigr)  .
\eeqn

We consider the spontaneous condensation of the $\rho$ meson fields in the background of the magnetic field~\eq{eq:Aext} in the vicinity of the transition temperature: $B >B_c$ with $B  - B_c \ll B_c$. One can show that in this case the $\rho$-meson condensate is very small, $|\rho| \ll m_\rho$, and the equations of motion for $\rho$-meson fields can be linearized~\cite{Chernodub:2010qx}. The smallness parameter is:
\beqn
\epsilon = \frac{g_s |\rho|_{\mathrm{max}}}{m_\rho} \simeq \sqrt{\frac{B  - B_c}{2 B_c}}, \qquad B  > B_c\,,
\label{eq:epsilon}
\eeqn
where $|\rho|_{\mathrm{max}} \equiv \max_{z} |\rho(z)|$ is the maximal value of the inhomogeneous condensate of the $\rho$ field.

The classical equations of motion in the magnetic field background are discussed in details in Ref.~\cite{Chernodub:2010qx}. The equation for the charged $\rho$--meson condensate can be written as follows:
\beqn
\DD\rho= 0\,,
\label{abrrhobar}
\eeqn
where the correction to this equation is of the order of $\mathcal{O}(\epsilon^4)$ with $\epsilon$ given in Eq.~\eq{eq:epsilon}. Basically, the solutions of Eq.~\eq{abrrhobar} reduce possible condensate solutions to the space of the lowest Landau levels. 

The neutral $\rho$--meson field $\rho^{(0)} \equiv \rho^{(0)}_x + i \rho^{(0)}_y$ can be expressed via the charged $\rho$--meson condensate in the following form:
\beqn
\rho^{(0)} = i \frac{g_s}{2}\frac{\partial}{- \Delta + m_0^2}\vert \rho \vert^2\,,
\label{rho0form}
\eeqn
where $\Delta = \partial\bar\partial \equiv \partial^2_x + \partial^2_y$ is the two--dimensional Laplacian,
\beqn
\frac{1}{- \Delta + m^2_0}(x_\perp) = \frac{1}{2 \pi} K_0(m|x_\perp|)\,,
\eeqn
is the two-dimensional Euclidean propagator of a scalar particle with the mass of the neutral vector meson~\eq{eq:m:rho0} and $K_0$ is a modified Bessel function.

Using the equations of motion for our model~\eq{eq:L:rho}, the mean energy density~\eq{dsgspot} can be expressed via a nonlocal function of the $\rho$--meson condensate $\rho$,
\beqn
\langle\mathcal{E}_\perp[\rho,\rho_0,A]\rangle & = & \frac{1}{2}B^2 + \langle\mathcal{E}_\perp[\rho]\rangle\,, 
\label{eq:E:full} \\
\langle\mathcal{E}_\perp[\rho]\rangle & = & \frac{e^2}{8}\langle\vert\rho\vert^2\rangle^2
+ \frac{1}{2}(m_\rho^2-eB)\langle\vert\rho\vert^2\rangle\nonumber \\
& & + \frac{g_s^2}{8} m_\rho^2 \Bigl\langle\vert\rho\vert^2\frac{1}{- \Delta + m_0^2}\vert\rho\vert^2 \Bigr\rangle\,,
\label{dsgspot3} 
\label{eq:E:cond}
\eeqn
where $\rho$ field is a minimum-energy solution of Eq.~\eq{abrrhobar} and the brackets $\langle \dots \rangle$ denote the average over the transverse plane:
\beqn
\langle \cO \rangle = \frac{1}{{\mathrm{Area}}_\perp} \int d x d y \, \cO(x,y)\,,
\label{eq:space:average}
\eeqn
and ${\mathrm{Area}}_\perp$ is the area of the transverse plane. 

The ground state solution of the $\rho$ condensate in the $B >B_c$ phase is given by the following expression~\cite{Chernodub:2011gs}, 
\beqn
\rho(\xx,\yy) & = & \sum_{n \in \Z} C_n h_n(\xx,\yy)\,,
\label{eq:rho:LLL}
\eeqn
where
\beqn
h_n(\xx,\yy) = \exp\left\{-\pi\frac{(\xx -n\nu L_B)^2 {-} i y (x - 2 n \nu L_B)}{L_B^2}\right\}\!, \qquad
\label{eq:h:n}
\eeqn
are the eigenstates corresponding to the Lowest Landau Level and 
\beqn
L_B=\sqrt{\frac{2\pi}{eB }}
\label{eq:LB}
\eeqn
is the magnetic length.

The parameters $C_n$ and $\nu$ should be chosen to minimize the energy~\eq{eq:E:cond} gained due to the condensation of the $\rho$--meson fields (the condensation energy), at a fixed value of magnetic field. The minimization of the energy functional is usually done by assuming that the (generally, complex) parameters obey the symmetry $C_n = C_{n+N}$ for an integer~$N$. 

The global minimum of the energy functional~\eq{dsgspot3} corresponds to a global minimum of the dimensionless quantity
\beqn
\beta_\rho = \left\langle \frac{|\rho^2|}{\langle |\rho^2| \rangle} \, \frac{m_0^2}{- \Delta + m_0^2} \frac{|\rho^2|}{\langle |\rho^2| \rangle} \right\rangle\,,
\label{eq:beta:rho}
\eeqn
which is an analogue of the Abrikosov ratio 
\beqn
\beta_A = \langle |\phi|^4 \rangle/\langle |\phi|^2 \rangle^2\,,
\label{eq:beta:A}
\eeqn
used in the Ginzburg-Landau theory of ordinary superconductors~\cite{Chernodub:2011gs}. Notice that $\beta_\rho \to \beta_A$ the limit $m_0 \to \infty$.

In Ref.~\cite{Chernodub:2011gs} we have found that the condensation energy density~\eq{eq:E:cond} -- with $\langle {\cal E} \rangle$ given in Eq.~\eq{dsgspot3} -- is minimized for $N=2$ 
with $C_0 = i C_1$ and 
\beqn
\nu = \frac{\sqrt[4]{3}}{\sqrt{2}} = 0.9306\dots\,. 
\label{eq:nu}
\eeqn
This choice of the parameters corresponds to an equilateral triangular (called also hexagonal) lattice with $N=2$, similarly to the case of the Abrikosov lattice in the Ginzburg--Landau model. All lattices with odd values of $N$ possess higher energies than the $N=2$ case while all even--$N$ lattices are reduced to the $N=2$ case. 

The parameters of Eq.~\eq{eq:rho:LLL} corresponding to the ground state are as follows:
\beqn
C_n = C_0 \, \alpha_n , \quad \alpha_{2\mathbb{Z}} = 1 , \quad  \alpha_{2\mathbb{Z}+1} = i \,.
\label{eq:coefficients:C}
\eeqn
The overall coefficient $C_0 = C_0(B )$ of the solution~\eq{eq:rho:LLL} and \eq{eq:coefficients:C} can then be calculated by a requirement of the minimization of the condensation energy~\eq{eq:E:cond} or the $\beta_\rho$ ratio~\eq{eq:beta:rho}. The mean-field behavior of the coefficient $C_0$ is calculated in Ref.~\cite{Chernodub:2011gs} and is also presented in Eq.~\eq{eq:C0}.

It was numerically found that in the ground state close to the critical magnetic field~\eq{eq:Bc} at $B = 1.01 B_c$ the ratio~\eq{eq:beta:rho} takes the following value~\cite{Chernodub:2011gs}:
\beqn
\beta_\rho (B = 1.01 B_c) = 1.0192 \dots \,.
\label{eq:beta:min}
\eeqn

\begin{figure} 
\begin{center}
\includegraphics[scale=0.5]{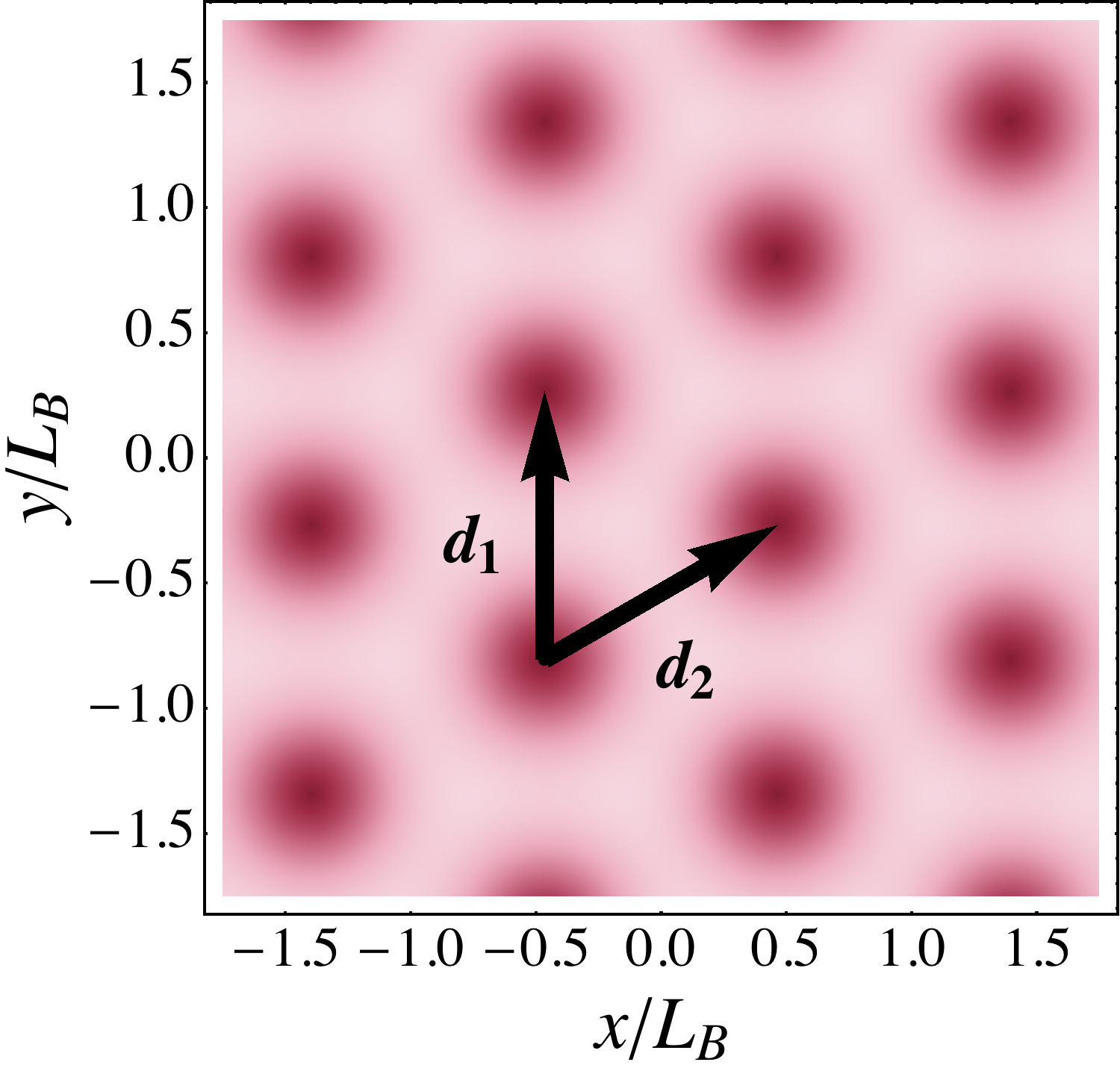}
\end{center}
\caption{The density plot of the $\rho$--meson condensate~\eq{eq:rho:LLL} in the ground state. The coordinates are given in terms of the magnetic length~\eq{eq:LB}. The darker regions corresponds to positions of the $\rho$ vortices where the superconducting density is small ($\rho \equiv 0$ at the centers of the vortices). In the mean-field ground state the vortices form the hexagonal lattice with the basic lattice vectors ${\bs d}_1$ and ${\bs d}_2$, Eq.~\eq{eq:basic:vectors}.}
\label{fig:rho:lattice}
\end{figure}

The $\rho$--meson condensate is an inhomogeneous function of the transverse coordinates $x$ and $y$ and is independent of the longitudinal $z$ coordinate~\eq{eq:rho:LLL}. The inhomogeneities are caused by the presence of an infinite periodic lattice of the so-called $\rho$ vortices which are parallel to the magnetic field. The $\rho$--vortex is a stringy topological defect in the $\rho$ condensate as the phase of the $\rho$ field winds by $2 \pi$ around the vortex center. There is one $\rho$ vortex per unit area $L_B^2 = 2 \pi / |e B|$ of the transverse plane. According to calculations in DSGS model~\cite{Chernodub:2011gs}, supported also by the holographic models~\cite{Bu:2012mq}, the vortices arrange themselves in the transverse plane in the form of a hexagonal (equilateral triangular) lattice, Fig.~\ref{fig:rho:lattice}. The hexagonal periodic lattice is encoded in the particular form of the parameters~$\alpha_n$ and $\nu$ in Eq.~\eq{eq:coefficients:C}.

For the hexagonal lattice solution~\eq{eq:rho:LLL} the basic lattice vectors are as follows:
\beqn
{\bs d}_1 = \frac{L_B}{\nu} \left(0, 1 \right)\,, \qquad 
{\bs d}_2 = \frac{L_B}{\nu} \left(\frac{\sqrt{3}}{2}, \frac{1}{2} \right)\,.
\label{eq:basic:vectors}
\eeqn

The solution~\eq{eq:rho:LLL} is periodic, up to a phase factor, with respect to the shifts ${\bs d}_1$ and ${\bs d}_2$:
\beqn
\rho({\bs x} + {\bs d}_a) & = & e^{i \pi {\bs x} \times {\bs d}_a/L_B^2} \rho({\bs x})\,, \qquad a = 1,2\,,
\label{eq:shifts}
\eeqn
where ${\bs a} \times {\bs b} = a_i \epsilon_{ij} b_j \equiv a_x b_y - a_y b_x$ is the vector product, and the latin superscript $a$ labels the basic vectors~\eq{eq:basic:vectors}.

\section{Crystal wavefunctions}
\label{sec:basis}

\subsection{Definition}

According to Eq.~\eq{eq:shifts}, the ground state solution~\eq{eq:rho:LLL}, \eq{eq:h:n}, \eq{eq:coefficients:C} is not invariant under the translational shifts along basic vectors~\eq{eq:basic:vectors} in the transverse plane. In order to study the vibrations of the vortex lattice it is convenient to describe the full basis of the eigenstates of the Lowest Landau Level~\eq{eq:h:n} by the so-called ``magnetically translated'' states. We follow the standard procedure which is used to study the vibrations of the vortex lattice in usual superconductors~\cite{ref:type-II:Review}.

We will perturb the vortex lattice both in transverse and longitudinal dimensions by slightly deformed configurations carrying so called ``crystal quasimomentum'' $k = ({\bs k},k_z)$ where 
${\bs k} = (k_x,k_y)$ and $k_z$ are the transverse and longitudinal quasimomenta, respectively. These quasimomentum states are space dependent functions, close to the original lattice configuration, but shifted to hold a relative momentum in a special way~\cite{ref:type-II:Review}:
\beqn
\begin{array}{rcl}
\rho_{k}(\bs x,z) & = & e^{i k_z z} \rho_{\bs k}(\bs x)\,, \\[1mm]
\rho_{\bs k}(\bs x) & = & e^{i\bs k\bs x}\psi_{\bs k}(\bs x)\,, 
\end{array}
\label{bloch}
\eeqn
where
\beqn
\psi_{\bs k}(\bs x) = \rho \Bigl({\bs x} + \frac{2 {\tilde{\bs k}}}{eB} \Bigr) \equiv \rho \Big({\bs x } +\frac{L_B^2}{\pi} {\tilde{\bs k}} \Big)\,,
\label{eq:psi}
\eeqn
and ${\tilde k}_i = \epsilon_{ij} k_j$ is the conjugate momentum. The shifted ground state function~$\psi_{\bs k}$ satisfies the classical equations of motion, in particular, Eq.~\eq{abrrhobar}. By construction, the momentum states~\eq{bloch} and \eq{eq:psi} satisfy the periodicity rule~\eq{eq:shifts} for arbitrary momentum $\bs k$:
\beqn
\rho_{\bs k}({\bs x} + {\bs d}_a) & = & e^{i \pi {\bs x} \times {\bs d}_a/L_B^2} \rho_{\bs k}({\bs x})\,, \qquad a = 1,2\,.
\label{eq:shifts:k}
\eeqn
The unperturbed ground--state function~\eq{eq:rho:LLL} corresponds to the function~\eq{bloch} with zero quasimomentum, $\rho \equiv \rho_0$.

It is convenient to define new dimensionless transversal coordinates:
\beqn
\bs x'= \frac{1}{L_B} {\bs x}\,,\qquad  
\bs k' =\frac{L_B}{\pi} {\bs k}\,, 
\label{k}
\eeqn
and then drop the primes, so that the the explicit form of the momentum eigenfunctions~\eq{bloch} is as follows:
\beqn
\rho_{\bs \kk}(\xx,\yy) & = & e^{i\pi (\xx\kk_x+\yy\kk_y)}\rho(\xx-\kk_x,\yy+\kk_y) \nonumber \\
& \equiv & C_0 e^{i\pi\left(\pi\xx\yy-\kk_x\kk_y\right) + 2\pi i\yy\kk_y } 
\label{eq:rho:k:explicit} \\
& & \cdot \sum_{n \in \Z} \alpha_n e^{-\pi(x-n\nu-\kk_y)^2+2\pi in\nu\yy+2\pi in\nu\kk_x}. \nonumber
\eeqn

In subsequent sections we will define the excited vortex states in terms of the quasimomentum wavefunctions~\eq{bloch}. These wavefunctions will eventually enter the perturbed energy density~\eq{dsgspot3}, which consists of two-- and four--point functions with respect to the average over the transverse space~\eq{eq:space:average}. For the sake of further convenience, in the rest of this section we define the two-- and four--point functions of the quasimomentum wavefunctions~\eq{bloch} in the transverse plane. The dependence of the wavefunctions $\rho_k = e^{i k_z z} \rho_{\bs k}({\bs x})$ on the longitudinal coordinate $z$ is omitted below since it gives rise to trivial phase factors only. 

\begin{figure} 
\begin{center}
\includegraphics[scale=0.3]{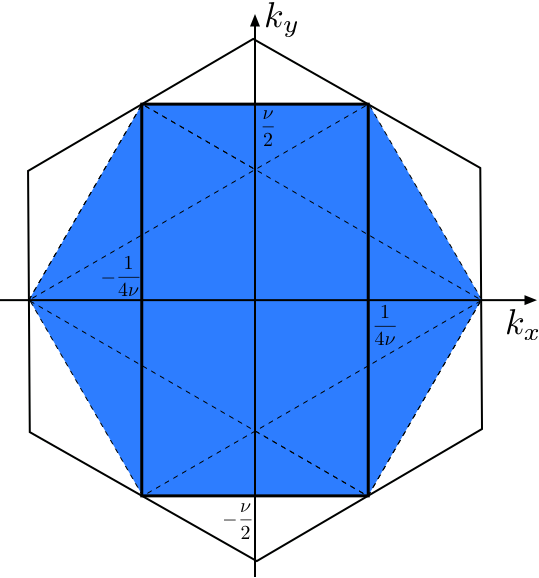}
\end{center}
\caption{Brillouin zone in the transverse momentum space (outer hexagon) containing the range where our low momentum calculation holds (inner solid rectangle). Symmetries extend the latter rectangle to two rotated copies (dashed rectangles), combining to the inner (blue) hexagon part of the Brillouin zone for which our findings are valid.
}
\label{fig:BZ}
\end{figure}

It is well known that the periodicity of the functions (\ref{eq:psi}) is inherited by the quasi-momenta, which can therefore be restricted to an elementary region in momentum space, the Brillouin-zone. This zone is then defined by the reciprocal vectors from $\bs d_a$ (\ref{eq:basic:vectors}):
\beqn
(\tilde{\bs d}_a)_i=\frac{eB}{2}\epsilon_{ij}(\bs d_a)_j\rightarrow \frac{1}{L_B}\epsilon_{ij}(\bs d_a)_j,
\eeqn
taking into account (\ref{k}).
While this periodicity allows a freedom in choosing the Brillouin zone, it is customary to choose it symmetrically around the origin and  such that it respects the symmetries of the original lattice. In this way one can clearly identify the small momenta corresponding to the infrared regime we're interested in. Since we are mainly interested in the infrared behavior of the phonon spectrum, in the remainder we will always assume all quasi-momenta sufficiently small, more explicitly:
\beqn
\kk_x,{l}_x \in \Big[-\frac{1}{4\nu},\frac{1}{4\nu}\Big], \quad \kk_y,{l}_y \in \Big[-\frac{\nu}{2},\frac{\nu}{2}\Big].
\label{bril}
\eeqn
Given the symmetries of the system, one can extend the region above to a small hexagon inside the Brillouin zone as shown in Fig.~\ref{fig:BZ}. We have not tested the validity of our calculations outside this range, nor have we extended the calculations to include the full Brillouin zone. Since our results coincide in the limiting case of Abrikosov lattices with the known result, we expect good agreement in the whole Brillouin zone. 

\subsection{Two-point function}

The two-point function of the transverse wavefunctions $\rho_{\bs k}({\bs x})$ can be calculated explicitly using Eq.~\eq{eq:rho:k:explicit}:
\beqn
\sbraket{\rho_{{\bs l}}^*\rho_{\bs \kk}} & = & |C_0|^2 \sum_{m,n \in \Z} \alpha_n\alpha_m^*e^{-\pi[ (\xx-n\nu-\kk_y)^2+(\xx-m\nu-{l}_y)^2 ]} \nonumber \\
& & \cdot e^{2i\pi\nu(n-m)y+2\pi i\nu (\kk_x n-{l}_x m) + 2\pi i(\kk_y-{l}_y)y}. \quad
\eeqn
By shifting the $x$ coordinate, $\xx \to \xx+n\nu+\kk_y$, we arrive to the following equation:
\beqn
\sbraket{\rho_{{\bs l}}^*\rho_{\bs \kk}} & = & \frac{L_B^2 |C_0|^2}{{\mathrm{Area}}_\perp}\int d\xx d\yy e^{-2\pi\xx^2} \sum_{n \in \Z} \vert\alpha_n\vert^2 e^{2\pi in\nu(\kk_x-{l}_x)} \ \quad
\nonumber \\
& = & \frac{|C_0|^2}{\sqrt{2} \nu} \delta(\bs \kk - {\bs l})\,.
\label{eq:two:point}
\eeqn

\subsection{Four-point function}

The (normalized) nonlocal correlation functions,
\beqn
Q_{{\bs l}_2, {\bs k}_2, {\bs l}_1, {\bs k}_1} = \frac{1}{|C_0|^4} \Big\langle \rho^*_{{\bs l}_2}\rho_{{\bs k}_2}^{~}\frac{M^2}{-\Delta+M^2}\rho^*_{{\bs l}_1}\rho_{{\bs k}_1^{~}} \Big\rangle\,,
\label{Qdef}
\eeqn
correspond to the last term in the transverse energy functional~\eq{dsgspot3}. Inserting Eq.~\eq{eq:rho:k:explicit} into Eq.~\eq{Qdef} and after an extensive algebra we find the following expression:
\beqn
Q_{{\bs l}_2, {\bs k}_2, {\bs l}_1, {\bs k}_1} & = & \frac{M^2}{2\nu^2}\sum_{m,n \in \Z} \frac{e^{-\pi\left(\bs c-\bs X_{m,n}\right)^2+2\pi i \epsilon_{ij}b^iX^j_{m,n}}}{4\pi^2\left(\bs c-\bs X_{m,n}\right)^2+M^2} \nonumber \\
& & \equiv {\cal Q}(\bs b,\bs c,M)\,,
\label{Q}
\eeqn
depending only on two linear combinations of the momenta $\bs b$ and $\bs c$ given below and the dimensionless mass 
\beqn
M = m_0 L_B\,,
\label{eq:M}
\eeqn
with $L_B$ is defined in Eq.~\eq{eq:LB}. The vectors $\bs b$, $\bs c$ and ${\bs X}_{m,n}$ in Eq.~\eq{Q} are as follows:
\beqn
\bs b & = & \frac{\bs k_{1}+\bs l_{1}-\bs k_{2}-\bs l_{2}}{2}\,,
\label{eq:b}\\
\bs c & = & \frac{\bs l_1-\bs k_1-\bs l_2+\bs k_2}{2}\,,
\label{eq:c}\\
\bs X_{m,n} &=& m\, \tilde {\bs d}_2+n\,\tilde {\bs d}_1\,\\
& =&m\,\nu^{-1}\Big(\frac{1}{2} \bs e_x - \nu^2 \bs e_y\Big)+n\,\nu^{-1}\bs e_x\,,
\label{eq:e}
\eeqn
where $\bs e_x$ and $\bs e_y$ are the unit vectors in the $x$ and $y$ directions, respectively. A detailed calculation of the correlation function~\eq{Q} is presented in Appendix~A.

Notice that in our gauge~\eq{eq:Aext} the function~\eq{Q} is always a real function, which is even in both arguments:
\beqn
Q_{{\bs l}_2, {\bs k}_2, {\bs l}_1, {\bs k}_1} \equiv Q_{{\bs k}_2, {\bs l}_2, {\bs k}_1, {\bs l}_1}\,, \quad {\mathrm{Im}}\, Q_{{\bs l}_2, {\bs k}_2, {\bs l}_1, {\bs k}_1} = 0\,.
\label{eq:Q:properties}
\eeqn

One can check Eq.~\eq{Q} by calculating the $\beta_\rho$ ratio~\eq{eq:beta:rho}:
\beqn
\beta_\rho \equiv 2\nu^2Q_{0,0,0,0} = \sum_{m,n \in \Z}
\frac{M^2 e^{-\pi  {\bs X}^2_{m,n}}}{4\pi^2 {\bs X}^2_{m,n} + M^2} \,.
\label{eq:beta:check}
\eeqn
The $\beta_\rho$ ratio depends on magnetic field $B$ via the dependence of the mass parameter $M = M(B)$ according to Eqs.~\eq{eq:LB} and \eq{eq:M}. An explicit calculation gives us that Eq.~\eq{eq:beta:check} at the critical magnetic field~\eq{eq:Bc} reproduces the known numerical value~\eq{eq:beta:min}. In the vicinity of the transition, $B \simeq B_c$, the parameter $\beta_\rho$ depends on the value of magnetic field $B$ very weakly. Parametrically, 
\beqn
\beta_\rho(B) = 1.01937 - 0.01702 \left(\frac{B}{B_c} - 1\right) + \dots\,,
\label{eq:beta:rho:weak}
\eeqn
for $|B - B_c|\ll B_c$. In the limit $M\to \infty$ we recover the standard result $\beta_A = 1.16$ for the value of the Abrikosov ratio for an equilateral triangular (hexagonal) lattice~\cite{ref:type-II:Review}.

Substituting the ground state wavefunction, Eq.~\eq{eq:rho:k:explicit} with ${\bs k} = 0$, into the energy functional~\eq{eq:E:cond} we compute, term by term, the ground-state energy:
\beqn
\mathcal{E}_\perp^{(0)} & = & \frac{e^2 \vert C_0\vert^4}{16 \nu^2} + \frac{m_\rho^2-eB}{2\sqrt{2}\nu}\vert C_0\vert^2 \nonumber \\
& & + \frac{g_s^2 m_\rho^2}{8 m_0^2} Q_{0,0,0,0}\vert C_0\vert^4\,,
\label{E0}
\eeqn
where we used the definition~\eq{Qdef}. 

The value of $\vert C_0\vert$ is then determined by the minimum of the energy~\eq{E0}. Given the phase degeneracy of the prefactor $C_0$ of the condensate $\rho_0$, we choose this prefactor to be a real number, $C_0 = |C_0|$. Then the ground state for $B > B_c$ is defined by Eqs.~\eq{eq:rho:LLL}, \eq{eq:h:n} and \eq{eq:coefficients:C} with the prefactor
\beqn
C_0 (B) & = & \sqrt{\frac{2\sqrt{2} (eB-m_\rho^2) \nu}{e^2+(g_s^2-e^2)\beta_\rho}} \label{eq:C0} \\
& \simeq &  0.2733 \sqrt{e B - e B_c} \simeq 0.1504 \sqrt{B - B_c}\,. \nonumber
\eeqn
where we took into account the first equality in Eq.~\eq{eq:beta:check} and Eq.~\eq{eq:m:rho0}. In numerical estimation of the prefactor~\eq{eq:C0} we used $\alpha_{\mathrm{em}} \equiv e^2/4\pi = 1/137$, the phenomenological value~\eq{eq:gs} for the coupling $g_s$, and the values~\eq{eq:nu} and \eq{eq:beta:min} for, respectively, the parameter $\nu$ and for the $\beta_\rho$ ratio in the ground state. 

Equation~\eq{E0} allows us to compute the energy of the ground state at $B > B_c$:
\beqn
& & \mathcal{E}_\perp^{(0)}(B) = - \frac{(e B - e B_c)^2}{2[e^2 + (g_s^2 - e^2) \beta_\rho]} 
\label{eq:E:cone:num} \\
& & \qquad \simeq - 0.0142 (e B - e B_c)^2 \simeq - 1.3 \times 10^{-3} (B - B_c)^2\,.
\nonumber
\eeqn
Obviously, in the low-$B$ phase the condensation energy is zero, $\mathcal{E}_\perp^{(0)}(B<B_c) = 0$.

\section{Dispersion relation for phonons}
\label{sec:phonons}

We perturb the mean-field solution for the ground state~\eq{eq:rho:k:explicit}, $\rho \equiv \rho_{\bs 0}({\bs x})$, by adding the states which carry a nonzero quasimomentum $\bs k$ in the transverse plane:
\beqn
\rho_{\mathrm{ph}}({\bs x}) = \sum_{\bs k}\, c_{\bs k} \rho_{\bs k} ({\bs x}), \qquad c_{|{\bs k}|\neq 0} \ll c_0 = 1\,.
\label{eq:rho:pert}
\eeqn
Our strategy is to substitute the perturbed wavefunction~\eq{eq:rho:pert} into the transverse energy functional~\eq{dsgspot3}, and expand the latter expression over the coefficients $c_{\bs k}$:
\beqn
\mathcal{E}_\perp[\rho_{\mathrm{ph}}] = \mathcal{E}_\perp^{(0)} + \mathcal{E}^{(2)}_\perp + O(c_{\bs{k}}^4)\,,
\eeqn
with $\mathcal{E}_\perp^{(0)}  \equiv \mathcal{E}_\perp[\rho_0]$ is the ground state energy~\eq{eq:E:cone:num} and $\mathcal{E}^{(2)}_\perp \sim c_{\bs{k}} c_{\bs{k}'}$ is the quadratic term of the phonon contribution to the energy. Then, the phonon eigenfunctions may be found by diagonalisation of the quadratic part $\mathcal{E}^{(2)}_\perp$ with respect to the coefficients $c_{\bs k}$. 

\subsection{Propagation in the transverse plane}

Due to the orthogonality of the magnetically translated wavefunctions~\eq{eq:two:point} the quadratic term of the energy functional~\eq{dsgspot3} is diagonal in the coefficients $c_{\bs k}$. However, the quartic term of this functional can, in general, mix four independent phonon modes of the perturbed wavefunction~\eq{eq:rho:pert}. Using the properties of the four-point function~\eq{eq:Q:properties}, one can show that number of independent mixing modes is reduced to a half of that upon complex conjugation.
The mixing of the remaining two modes can be described by the following two-component vector,
\beqn
v_{\bs k} = (c_{\bs k},c^*_{-{\bs k}})^T\,,
\eeqn
and the mixing term can be written as follows:
\beqn
\braket{\rho^*\rho\frac{M^2}{-\Delta+M^2}\rho^*\rho}  = 
\frac{1}{2} \sum_{{\bs k}} v^\dag_{\bs k} \cdot {\hat Q} \cdot v_{\bs k}\,,
\eeqn
with the matrix
\beqn
{\hat Q} = 2 \left(\begin{array}{cc} Q_{{\bs k},{\bs k},0,0}+ Q_{{\bs k},0,0,{\bs k}} & Q_{{\bs k},0,-{\bs k},0}^* \\  
Q_{{\bs k},0,-{\bs k},0} & Q_{{\bs k},{\bs k},0,0}+ Q_{{\bs k},0,0,{\bs k}} \end{array}\right). \qquad
\label{eq:hat:Q}
\eeqn

The eigenvalues of the matrix~\eq{eq:hat:Q} are as follows:
\beqn
\lambda_{{\bs k},\pm} & = & 2 \left(Q_{{\bs k},{\bs k},0,0}+ Q_{{\bs k},0,0,{\bs k}}\pm\vert Q_{{\bs k},0,-{\bs k},0}\vert\right) \nonumber \\
& \equiv & 2\left({\cal Q}({\bs k},0,M)+{\cal Q}(0,\bs k,M)\pm |{\cal Q}(\bs k,\bs k,M)|\right), \qquad 
\label{lambda}
\eeqn
Notice that 
\beqn
Q_{0,0,0,0}=\frac{\lambda_{0,+}}{6}=\frac{\lambda_{0,-}}{2} \equiv \frac{\beta_\rho}{2\, \nu^2} \,.
\label{beta}
\eeqn

The eigenvectors of the matrix~\eq{eq:hat:Q},
\beqn
o_{\bs k} = \frac{c_{\bs k}+c_{-{\bs k}}^*}{2}\,,
\qquad 
a_{\bs k}=\frac{c_{\bs k} - c^*_{-{\bs k}}}{2i}\,,
\eeqn
define the optical (massive) and acoustic (massless) phonon modes with the eigenvalues $\lambda_{{\bs k},+}$ and $\lambda_{{\bs k},-}$, respectively.

To the lowest order the energy functional of the optical and acoustic transverse modes reads as follows:
\beqn
\mathcal{E}^{(2)}_\perp = \sum_{\bs k} \mathcal{E}^{(2)}_\perp ({\bs k})\,,
\eeqn
where the contributions from the individual modes are diagonal in the coefficients $c_{\bs k}$:
\begin{widetext}
\beqn
\mathcal{E}^{(2)}_\perp ({\bs k}) & = & c_{\bs k}^*c_{\bs k} \left(\frac{e^2}{4} \langle \vert\rho_0\vert^2 \rangle \langle\rho_{\bs k}^*\rho_{\bs k} \rangle 
+ \frac{1}{2}(m_\rho^2-eB)\langle \rho_{\bs k}^*\rho_{\bs k} \rangle \right) 
+ \frac{g_s^2 m_\rho^2 |C_0|^2 }{4 \, m_0^2} \left( Q_{{\bs k},{\bs k},0,0} + Q_{{\bs k},0,0,{\bs k}}\right)c_{\bs k} c^*_{\bs k} \nonumber \\
& & + \frac{g_s^2 m_\rho^2 |C_0|^2}{8}  c_{\bs k} c_{-{\bs k}} Q_{0,{\bs k},0,-{\bs k}}+\frac{g_s^2 m_\rho^2 |C_0|^2}{8} {c_{\bs k}^*}c_{-{\bs k}}^* Q_{{\bs k},0,-{\bs k},0} \\
& = & o_{\bs k}^2 \cdot \left(\frac{e^2}{8\nu^2}\vert C_0\vert^2+\frac{m_\rho^2-eB}{\sqrt{8}\nu}+\frac{g_s^2}{8}\frac{m_\rho^2}{m_0^2}\vert C_0\vert^2\lambda_{{\bs k},+}\right)
+ a_{\bs k}^2 \cdot \left(\frac{e^2}{8\nu^2}\vert C_0\vert^2+\frac{m_\rho^2-eB}{\sqrt{8}\nu}+\frac{g_s^2}{8}\frac{m_\rho^2}{m_0^2}\vert C_0\vert^2\lambda_{{\bs k},-}\right).
\label{eq:E1}
\nonumber
\eeqn
\end{widetext}

The above expression can be simplified further using Eq.~\eq{beta} and the fact that the parameter $C_0$, given explicitly in Eq.~\eq{eq:C0}, corresponds to the minimum of energy~\eq{E0}:
\beqn
\mathcal{E}^{(2)}_\perp ({\bs k}) & = & \mathcal{E}^{(2)}_{\perp,O} ({\bs k}) + \mathcal{E}^{(2)}_{\perp,A} ({\bs k}) = \frac{g^2_s-e^2}{8}\vert C_0\vert^2 \nonumber \\
& & \cdot \Big[o_{\bs k}^2(\lambda_{{\bs k},+}-\lambda_{0,-})+a_{\bs k}^2(\lambda_{{\bs k},-}-\lambda_{0,-}) \Big]\,. \quad
\label{eq:E}
\eeqn

The phonon contribution to the transverse energy~\eq{eq:E} has a massless mode in the $a_{\bs k}$ spectrum -- the last term in Eq.~\eq{eq:E} vanishes if ${\bs k} = 0$ -- which are therefore called the ``acoustic'' modes. The $o_{\bs k}$ modes represent the massive ``optical'' modes. Below we concentrate on massless acoustic modes.

According to Eqs.~(\ref{eq:E}) and (\ref{beta}) the contribution of the acoustic modes to the energy can be rewritten as follows:
\beqn
\mathcal{E}^{(2)}_{\perp,A} ({\bs k}) = a_{\bs k}^2\frac{g^2_s-e^2}{8}\vert C_0\vert^2 \Big(\lambda_{{\bs k},-} - \frac{\beta_\rho}{\nu^2} \Big)\,,
\label{eq:E:phonons:trans}
\eeqn
with $C_0$ given in Eq.~\eq{eq:C0}. The eigenvalue $\lambda_{{\bs k},-}$ is explicitly given by Eqs.~(\ref{lambda}) and (\ref{Q}):
\beqn
& & \lambda_-({\bs k}) \equiv \lambda_{\bs k,-} = \frac{M^2}{2\nu^2} \Biggl( \sum_{\bs X} \frac{e^{-\pi(\bs k-\bs X)^2}}{4\pi^2 (\bs k-\bs X)^2+M^2}  
\label{eq:lambda:k}\\
& & 
+ \sum_{\bs X} \frac{e^{-\pi\bs X^2+2\pi i \bs k\times\bs X}}{4\pi^2 \bs X^2+M^2}- \left| \sum_{\bs X} \frac{e^{-\pi(\bs k-\bs X)^2+2\pi i \bs k\times\bs X}}{4\pi^2 (\bs k-\bs X)^2+M^2} \right| \Biggr).
\nonumber
\eeqn
and $\bs X = {\bs X}_{m,n}$ as given in Eq.~\eq{eq:e} with $m,n \in \Z$ running over the whole lattice. 

\subsection{Acoustic spectrum}

The phonon contribution~\eq{eq:E:phonons:trans} to the energy is, in fact, a potential (time-independent) energy of essentially transverse modes with zero longitudinal momentum, $k_z = 0$. The propagation in the longitudinal direction $z$ and the time evolution of the acoustic modes can naturally be taken into account by the following expression for the phonon fluctuations:
\beqn
\rho_{\mathrm{ph}}({\bs x},z,t) = \sum_{{\bs k},k_z}\, c_{{\bs k},k_z} e^{- i \omega_k t + i k_z z} \rho_{\bs k} ({\bs x})\,,
\label{eq:rho:pert:full}
\eeqn
where $k=({\bs k},k_z)$ according  to Eq.~\eq{bloch}.

The easiest way to obtain the dispersion relation for the phonon modes is to notice that the quadratic contribution to the time-dependent part of the Lagrangian~\eq{eq:L:rho}
coming from the phonon fluctuations~\eq{eq:rho:pert:full} is as follows:
\beqn
T & = & \sbraket{\mathcal{L}_t} =\frac{1}{4}\sbraket{\vert \partial_t\bar \rho\vert^2}+\frac{1}{2}\sbraket{\vert\partial_t\rho^{(0)}\vert^2} \nonumber \\
& = & \frac{\omega_k^2 a_k^2}{4}\sbraket{\vert \rho_k\vert^2}  = \frac{\omega_k^2 a_k^2}{4\sqrt{2}\nu}\,. 
\label{eq:T}
\eeqn
The expression~\eq{eq:T} plays a role of the kinetic energy of the phonon fluctuations. The potential energy is given by $V = \mathcal{E}^{(2)}_\perp ({\bs k}) + \mathcal{E}^{(2)}_{\|,A} (k_z)$, where 
\beqn
\mathcal{E}^{(2)}_{\|,A} (k_z) = \frac{a_k^2}{4\sqrt{2}\nu} k_z^2\,,
\label{eq:E:phonons:parallel}
\eeqn
is the contribution from the longitudinal phonons which can be obtained by substituting the wavefunction of phonon fluctuations~\eq{eq:rho:pert:full} into the expression for the full energy~\eq{dsgspot}. Since the energy is diagonal in the longitudinal wavefunctions, there is no term which mixes the longitudinal modes with different $k_\|$.

Solving the equations of motion is then equivalent to putting $T-V=0$. Then we obtain the following dispersion relation for the acoustic phonon modes:
\beqn
\omega_{k}^2 = \frac{2e (g_s^2-e^2)(B - B_c)}{e^2 + (g_s^2-e^2) \beta_\rho} \left[ \nu^2 \lambda_{-}\!\! \left( \frac{L_B {\bs k}}{\pi} \right) {-} \beta_\rho \right] \! {+} k_z^2, \qquad
\label{eq:omega:k:full} 
\eeqn
where the explicit form of $\lambda_-$ is given in Eq.~\eq{eq:lambda:k}. In deriving Eq.~\eq{eq:omega:k:full}  we have used Eqs.~\eq{eq:C0} and \eq{eq:E:phonons:trans}.

\begin{figure} 
\begin{center}
\includegraphics[scale=0.4]{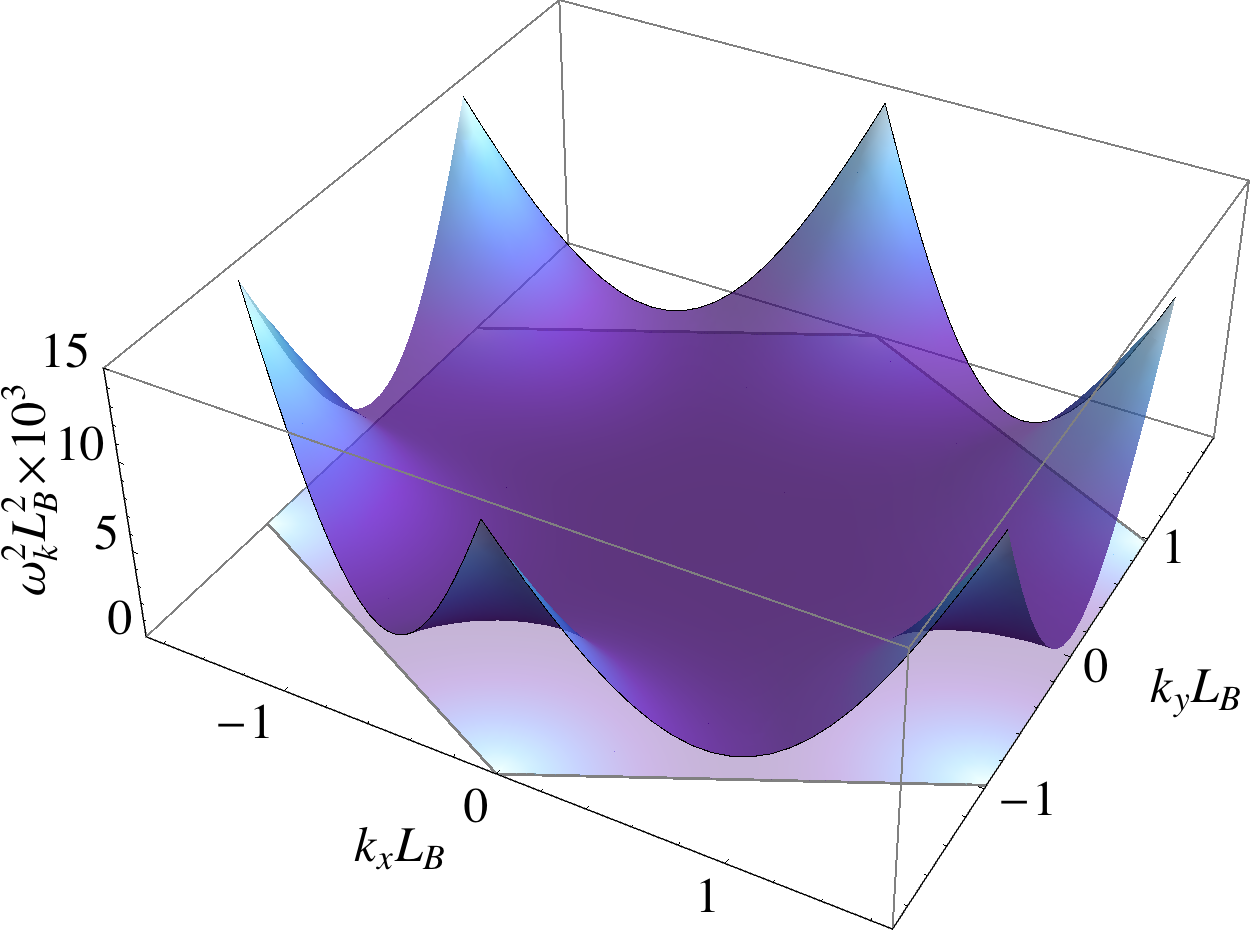}
\end{center}
\caption{The dispersion $\omega_k$ of the gapless (photon) modes in the transverse momentum space, ${\bs k} = (k_x,k_y)$ at $B = 1.01 B_c$. The projection to $(k_x,k_y)$ plane corresponds to the Brillouin zone shown in Fig.~\ref{fig:BZ}.}
\label{fig:omega2}
\end{figure}

The dispersion relation~\eq{eq:omega:k:full} for the transverse (with $k_z = 0$) acoustic modes in the Brillouin zone is shown in Fig.~\ref{fig:omega2}.  Qualitatively, the phonon spectrum in the $\rho$--vortex lattice is very similar to the spectrum of  phonons in Abrikosov vortex lattices of conventional superconductors~\cite{ref:type-II:Review}.

\subsection{Low-energy spectrum of acoustic phonons}

To obtain the infrared spectrum of the low-energy acoustic photons we expand the eigenvalue~\eq{eq:lambda:k} over small transverse momenta\footnote{We remind that according to Eq.~\eq{k} the dimensionless momentum $\bs k$ is related to the physical one a follows: $\bs k= L_B \bs k_{\mathrm{phys}}/ \pi$.} ${\bs k}^2$:
\beqn
& & \lambda_{-} (\bs k) = \frac{\beta_\rho}{\nu^2} + \alpha(B) \left({\bs k}^2\right)^2 + O\left(({\bs k}^2)^3\right)\,,  
\label{eq:lambda:IR} \\
& & \alpha (B = 1.01 B_c) \simeq 0.662\,.
\label{eq:alpha:B}
\eeqn
Similarly to the $\beta_\rho$ ratio~\eq{eq:beta:rho:weak}, the prefactor~\eq{eq:alpha:B} is practically insensitive to the magnetic field $B$ in the transition region.

In the case of the Abrikosov vortex lattice in a type--II superconductor -- which is achieved by taking the limit $M \to \infty$ in Eq.~\eq{eq:lambda:k} -- the coefficient in front of the quartic term in Eq.~\eq{eq:lambda:IR} is 4 times bigger: $\alpha_A \simeq 2.72$. 

Then we substitute the expansion~\eq{eq:lambda:IR} to the full phonon spectrum~\eq{eq:omega:k:full}, and get the following infrared spectrum of the acoustic phonons:
\beqn
\omega_k^2 & = & k_z^2 + f(B) \left({\bs k}^2\right)^2 + \dots
\,, 
\label{eq:omega:k:low}  \\
f(B) & = & \frac{C_f}{|eB|} \left(1 - \frac{B_c}{B} \right) + \dots
\,,
\label{eq:f:B}
\eeqn
where $C_f \simeq 0.455$ and the higher-order corrections in ${\bs k}_\perp^2$ and in $B - B_c$ are shown by the ellipsis. This expansion is valid in the vicinity of the transition $B \geqslant B_c$.

The longitudinal part of the phonon spectrum~\eq{eq:omega:k:low} contains massless (``soft'') phonon modes with the linear dispersion relation $\omega_{{\bs k} = 0, k_z} = k_z$. Thus, the acoustic phonons travel along the magnetic field axis with the speed of light and belong to type-I Nambu-Goldstone (NG) bosons. 

The transverse part of the spectrum~\eq{eq:omega:k:low} corresponds to the ``supersoft'' phonon modes which are described by the quadratic dispersion relation 
$\omega_{{\bs k}, k_z = 0} = \sqrt{f(B)} {\bs k}^2$. This mode corresponds to a type-II NG boson. Similar quadratic dispersion relations were found for kaon condensation in the color-flavor locked phase of QCD at a nonzero strange chemical potential~\cite{Schafer:2001bq,Miransky:2001tw}. Generalizations of the NG~\cite{Amado:2013xya} and pseudo-NG~\cite{Filev:2009xp} modes were found recently in the holographic approaches.

The purely transverse acoustic modes propagate with the following velocity (in units of speed of light $c$):
\beqn
v_\perp({\bs k}, k_z=0) = 2 \sqrt{f(B)} |{\bs k}|\,.
\eeqn
For example, at $B = 1.01 B_c$ a transverse acoustic phonon carrying energy $\omega_{{\bs k},0} = 1$\,MeV should travel with the velocity equal to 2\% of speed of light.

The dispersion relation for the low-energy phonons~\eq{eq:omega:k:low} in the $\rho$-vortex lattices has the same qualitative form as the dispersion relation for the acoustic phonons in Abrikosov vortex lattices in conventional superconductors. As in the case of usual superconductors, the ``super-softness'' of the transverse phonon mode may lead to an instability of the $\rho$ vortex lattice against thermal and quantum fluctuations. Depending on the strength of these fluctuations, the superconductor's vortex lattice may either withstand the perturbations or melt into either a vortex liquid~\cite{ref:Moore} or a vortex glass~\cite{ref:Fischer} (we refer a reader to the review~\cite{ref:type-II:Review} for further details). The study of stability of the vortex lattice is beyond the scope of this article.

\section{Conclusions}

We have demonstrated the existence of the acoustic phonon modes in a suggested superconducting phase of QCD vacuum at strong magnetic field. 

In the mean--field approach, the superconducting ground state resembles an Abrikosov vortex lattice in the mixed phase an ordinary type--II superconductor. The vortices are embedded in a vector quark-antiquark condensate with carries the quantum numbers of $\rho$ mesons. The vortices form a hexagonal lattice in the transverse plane with respect to the axis of the magnetic field. The lattice breaks translational and rotational symmetries of the coordinate space and leads to the appearance of the Nambu-Goldstone modes (acoustic phonons). 

We have shown that the acoustic vibrations of the vortex lattice along the direction of the magnetic field is a linear function of momentum. A phonon propagating in the transverse plane possesses a quadratic (supersoft) dispersion relation in the limit of small momenta. The longitudinal phonons propagate with the speed of light while the velocity of their transverse counterparts depends on their energy and it may be much smaller than the speed of light. In the infrared limit the spectrum of acoustic phonons is given by Eq.~\eq{eq:omega:k:low}.

The presence of the supersoft phonon modes is known to be crucial for the stability of the vortex lattice since these modes make an infrared divergent contribution to the free energy of the system~\cite{ref:type-II:Review}. As a result, the vortex lattice may become unstable and melt into a vortex liquid. 
In our paper we have found the supersoft mode in the phonon spectrum. This finding is in agreement with the results of the numerical simulations of lattice QCD in strong magnetic field background~\cite{Braguta:2013uc} which indicate the presence of the a liquid vortex state rather then an ordered vortex lattice.

\acknowledgments

The work of M.N.C. was partially supported by grant No. ANR-10-JCJC-0408 HYPERMAG of Agence Nationale de la Recherche (France). The work of J.V.D. was supported by a grant from La R\'egion Centre (France).

\appendix
\section{Explicit calculation of the quartic term}

In this Appendix we evaluate the four--point function~\eq{Qdef} which has the following explicit form:
\beqn
Q_{{\bs l}_2, {\bs k}_2, {\bs l}_1, {\bs k}_1} & = &
\iint\frac{dxdy}{L_xL_y} \iint\frac{dx' dy' }{L_xL_y} 
\nonumber \\
& & \prod_{i=1}^2\sum_{m_i \in \Z}\sum_{n_i \in \Z} M^2\delta (x- x')  f(y)  \qquad
\label{Q2} \\
& & \hskip -21mm \frac{\alpha h }{-\partial_x^2+4\pi^2 [\kk_{1,y}-{l}_{1,y}+\nu(n_1-m_1)]^2+M^2} g(x, x') \,,
\nonumber 
\eeqn
where $x'$ is an arbitrary coordinate and
\beqn
f(y) & = & e^{2\pi iy(k_{1,y}+k_{2,y}-l_{1,y}-l_{2,y})} \nonumber \\
& & \cdot e^{2\pi iy\nu (n_1+n_2-m_1-m_2)}, \nonumber\\
g(x, x') & = & e^{-\pi(x' - \nu n_2-k_{2,y})^2-\pi(x' - \nu m_2-l_{2,y})^2} 
\label{eq:fgh} \\
& & \cdot \, e^{-\pi(x-\nu n_1-k_{2,y})^2-\pi (x-\nu m_2-l_{2,y})^2}, \nonumber \\
h & = & e^{2\pi i\nu(\kk_{1,x}n_1+\kk_{2,x}n_2-{l}_{1,x}m_1-{l}_{2,x}m_2)}, \nonumber \\
\alpha & = & \alpha_{m_2}^*\alpha_{m_1}^*\alpha_{n_2}\alpha_{n_1}. \nonumber
\eeqn
We have assumed that we are working in a large but finite transverse space with dimensions $L_x \times L_y$. The derivative in Eq.~\eq{Q2} acts on the function $g(x, x')$ only. For the sake of brevity, we do not show explicitly all arguments of the functions $f$, $g$, $h$ and $\alpha$ which are defined in Eq.~\eq{eq:fgh}.

The integral in Eq.~\eq{Q2} over the $y$ coordinate can be taken as follows. Firstly, one gets:
\beqn
\int dy f(y) & = & \delta\bigl(k_{1,y}+k_{2,y} - l_{1,y}-l_{2,y}\nonumber \\
& & +\nu (n_1+n_2-m_1-m_2)\bigr)\,.
\label{f1}
\eeqn
Because of the constraint we put on momenta in the $y$ direction~\eq{bril}, one finds that the absolute value of the sum $k_{1,y}+k_{2,y} - l_{1,y}-l_{2,y}$ in Eq.~\eq{f1} should always be smaller than $2\nu$. Moreover, we need only the expressions with two nonzero momenta, so that the above sum turns out to be smaller than $\nu$. The later automatically implies $n_1+n_2-m_1-m_2 = 0$, so that 
\beqn
\int dy f(y) & = & \delta\left(k_{1,y}+k_{2,y}-l_{1,y}-l_{2,y}\right) \nonumber\\
& & \cdot \, \delta_{n_1+n_2-m_1-m_2}.
\label{dy}
\eeqn

The function $g(x,x')$ in Eq.~\eq{eq:fgh}  can be rewritten as follows: 
\beqn
g(x, x') & = & e^{-\pi[(x + x' - 2r \nu-a_y)^2+(x - x' - 2s\nu-b_y)^2]} \nonumber \\
& & \cdot \,  e^{- \pi [(2t\nu+c_y)^2+(2 u\nu+d_y)^2]}\,,
\eeqn
with
\al{
r&=\frac{n_1+m_1+n_2+m_2}{2}\,,&\bs a&=\frac{\bs k_{1}+\bs l_{1}+ \bs k_{2}+\bs l_{2}}{2}\,, \nonumber \\
s&=\frac{n_1+m_1-n_2-m_2}{2}\,,&\bs b&=\frac{\bs k_{1}+\bs l_{1}-\bs k_{2}-\bs l_{2}}{2}\,, \nonumber \\
t&=\frac{n_1-m_1-n_2+m_2}{2}\,,&\bs c&=\frac{\bs l_1-\bs k_1-\bs l_2+\bs k_2}{2}\,, \nonumber \\
u&=\frac{n_1-m_1+n_2-m_2}{2}\,,&\bs d&=\frac{\bs l_1-\bs k_1+\bs l_2-\bs k_2}{2}\,. \nonumber
}

Equation~(\ref{dy}) implies $u=d_y=0$. In terms of the new variables, we find $t=n_1-m_1$, $c_y=l_{1,y}-k_{1,y}$ and
\beqn
h=e^{2\pi i\nu(ua_x+tb_x-rd_x-sc_x)}\,.
\eeqn

To reproduce all possible integer values of $n_1,\ldots m_2$ we use the following summation formula:
\beqn
\tilde {\sum_{r,s,t}} \equiv \sum_{s,t\in \mathbb{Z}}\left(\sum_{r\in 2\mathbb{Z}+(s+t){\rm mod}2}\right)\,,
\eeqn
which takes into account the restriction $u=0$. An analysis of the $\alpha$--factor in Eq.~\eq{eq:fgh} under the same conditions reveals that
$\alpha=(-1)^{st}$. 

We also rewrite the non-local operator in (\ref{Q2}) as follows:
\beqn
& & \frac{1}{-\partial_x^2+4(\kk_{1,y}-{l}_{1,y}+\pi\nu(n_1-m_1))^2+M^2} \delta(x - x')\nonumber \\
& = & \int dq\frac{e^{2\pi iq(x-\tilde x)}}{4\pi^2q^2+4\pi^2(\nu t-c_y)^2+M^2}\,.
\eeqn
 Changing integration variables to 
\beqn
u=x+\tilde x,\quad v= x'-x,\quad dudv=2dxd x',
\eeqn
and  using all the above ingredients, the expression (\ref{Q2}) becomes:
\begin{widetext}
\beqn
Q_{l_2,k_2,l_1,k_1} & = & M^2\frac{ \delta(d_y)}{L_xL_y}\int dudvdq \tilde{\sum_{r,s,t}}(-1)^{st} e^{2\pi i[\nu(tb_x-rd_x-sc_x)+qv]} 
\frac{e^{-\pi\left((u-r \nu-a_y)^2+(v-s\nu-b_y)^2+(t\nu-c_y)^2\right)}}{q^2+4\pi^2(\nu t-c_y)^2+M^2} \nonumber\\
& = & M^2\frac{\delta(d_y)}{L_x L_y}\int q \tilde{\sum_{r,s,t}}\mathbf{\alpha}
e^{-2\pi i\nu(sc_x-sq+rd_x)} \frac{e^{-\pi q^2-\pi(t\nu-c_y)^2+2\pi i(\nu tb_x+qb_y)}}{q^2+4\pi^2(\nu t-c_y)^2+M^2}\,.
\eeqn
\end{widetext}
We will now take the  $r$ summation and constrain $d_x$ to twice the region (\ref{bril}) as we will need at most two non-zero quasimomenta, i.e.  $d_x$ should be smaller than $\frac{1}{2\nu}$:
\al{
\sum_{r\in 2\mathbb{Z}} e^{- 2\pi i\nu rd_x}&=\frac{1}{2\nu}\sum_k\delta( d_x-\frac{1}{2\nu} k) \nonumber\\
\sum_{r\in 2\mathbb{Z}+1} e^{- 2\pi i\nu rd_x}&=\frac{1}{2\nu}\sum_k(-1)^k\delta( d_x-\frac{1}{2\nu} k) \nonumber\\
\tilde{\sum_{s,t,r}}(-1)^{st} e^{- 2\pi i\nu rd_x}&=\frac{1}{2\nu}\sum_{r,s,t}(-1)^{r(s+t)+st}\delta( d_x-\frac{1}{2\nu} r) \nonumber \\
&=\frac{1}{2\nu}\sum_{s,t}(-1)^{st}\delta(d_x) \,. \nonumber
}
The $s$ summation gives likewise
\al{
\sum_se^{-2\pi is\nu (c_x-q-\frac{t}{2\nu})}&=\sum_s\frac{1}{\nu}\delta(q- c_x-\frac{1}{\nu}(s-\frac{1}{2}t))\,. \nonumber
}
Using the identity $\delta(\bs d\,)\equiv L_xL_y$ we get:
\beqn
& & Q_{l_2,k_2,l_1,k_1} = \frac{M^2}{2\nu^2} \sum_{s,t} \nonumber \\
& & 
\frac{e^{-\pi\left(c_x-\frac{2s-t}{2\nu}\right)^2-\pi(t\nu-c_y)^2+2\pi i\left(\nu t b_x - b_y\frac{2s-t}{2\nu}\right)}}{4\pi^2\left(c_x-\frac{2s-t}{2\nu}\right)^2+4\pi^2(\nu t-c_y)^2+M^2}\,,
\qquad
\label{eq:Q:preliminary}
\eeqn
where we have also used the following property:
\beqn
e^{-i\pi(k_{1,x}k_{1,y}+k_{2,x}k_{2,y}-l_{1,x}l_{1,y}-l_{2,x}l_{2,y})} \delta({\bs d}) = e^{-2\pi i b_yc_x} \delta({\bs d})\,.
\nonumber
\eeqn
Redefining $t \to - t$ in Eq.~\eq{eq:Q:preliminary} we get the desired expression~\eq{Q}.


\begin{thebibliography}{99}

\bibitem{ref:magnetic:catalysis}
  I.~A.~Shovkovy,
  %``Magnetic Catalysis: A Review,''
  Lect.\ Notes Phys.\  {\bf 871}, 13 (2013);
  %%CITATION = ARXIV:1207.5081;%%
  H.~Suganuma and T.~Tatsumi,
  %``On The Behavior Of Symmetry And Phase Transitions In A Strong Electromagnetic Field,''
  Annals Phys.\  {\bf 208}, 470 (1991);
  %%CITATION = APNYA,208,470;%%
  %67 citations counted in INSPIRE as of 28 Feb 2013
  K.~G.~Klimenko, Z.\ Phys.\ C {\bf 54}, 323 (1992);
  %%CITATION = ZEPYA,C54,323;%%
  V.~P.~Gusynin, V.~A.~Miransky, I.~A.~Shovkovy,
  Phys.\ Rev.\ Lett.\  {\bf 73}, 3499 (1994);
  %%CITATION = HEP-PH/9405262;%%
  Phys.\ Lett.\ B {\bf 349}, 477 (1995);
  %%CITATION = HEP-PH/9412257;%%
  Nucl.\ Phys.\  {\bf B462}, 249 (1996).
  %%CITATION = HEP-PH/9509320;%% 

\bibitem{ref:Tc:rising}
  N.~O.~Agasian and S.~M.~Fedorov,
  %``Quark-hadron phase transition in a magnetic field,''
  Phys.\ Lett.\ B {\bf 663}, 445 (2008);
  %%CITATION = ARXIV:0803.3156;%%
 R.~Gatto, M.~Ruggieri,
  Phys.\ Rev.\ D {\bf 82}, 054027 (2010);
  %%CITATION = ARXIV:1007.0790;%%
  Phys.\ Rev.\  {D \bf 83}, 034016 (2011);
 %%CITATION = ARXIV:1012.1291;%%
  A.~J.~Mizher, M.~N.~Chernodub and E.~S.~Fraga,
  %``Phase diagram of hot QCD in an external magnetic field: possible splitting of deconfinement and chiral transitions,''
  Phys.\ Rev.\ D {\bf 82}, 105016 (2010);
  %%CITATION = ARXIV:1004.2712;%%
    V.~Skokov,
  %``Phase diagram in an external magnetic field beyond a mean-field approximation,''
  Phys.\ Rev.\ D {\bf 85}, 034026 (2012);
  %%CITATION = ARXIV:1112.5137;%%
  V.~D.~Orlovsky and Y.~A.~Simonov,
  %``The quark-hadron thermodynamics in magnetic field,''
  arXiv:1311.1087 [hep-ph];
  %%CITATION = ARXIV:1311.1087;%%
  J.~O.~Andersen, W.~R.~Naylor and A.~Tranberg,
  %``Chiral and deconfinement transitions in a magnetic background using the functional renormalization group with the Polyakov loop,''
  arXiv:1311.2093 [hep-ph];
  %%CITATION = ARXIV:1311.2093;%%
  E.~S.~Fraga, B.~W.~Mintz and J.~Schaffner-Bielich,
  %``A search for inverse magnetic catalysis in thermal quark-meson models,''
  arXiv:1311.3964 [hep-ph].
  %%CITATION = ARXIV:1311.3964;%%
  
\bibitem{ref:lattice:results}  
  G.~S.~Bali, F.~Bruckmann, G.~Endrodi, Z.~Fodor, S.~D.~Katz, S.~Krieg, A.~Schafer and K.~K.~Szabo,
  %``The QCD phase diagram for external magnetic fields,''
  JHEP {\bf 1202}, 044 (2012);
  %%CITATION = ARXIV:1111.4956;%%
    F.~Bruckmann, G.~Endrodi and T.~G.~Kovacs,
  %``Inverse magnetic catalysis and the Polyakov loop,''
  JHEP {\bf 1304}, 112 (2013).
  %%CITATION = ARXIV:1303.3972;%%
  
\bibitem{ref:magnetic:matter}
  F.~Preis, A.~Rebhan and A.~Schmitt,
  %``Inverse magnetic catalysis in dense holographic matter,''
  JHEP {\bf 1103}, 033 (2011);
  %%CITATION = ARXIV:1012.4785;%%
%  F.~Preis, A.~Rebhan and A.~Schmitt,
  %``Holographic baryonic matter in a background magnetic field,''
  J.\ Phys.\ G {\bf 39}, 054006 (2012);
  %%CITATION = ARXIV:1109.6904;%%
  V.~Dexheimer, R.~Negreiros and S.~Schramm,
  %``Hybrid Stars in a Strong Magnetic Field,''
  Eur.\ Phys.\ J.\ A {\bf 48}, 189 (2012);
  %%CITATION = ARXIV:1108.4479;%%
  X.~Kang, M.~Jin, J.~Xiong and J.~Li,
  %``The influence of magnetic field on the pion superfluidity and phase structure in the NJL model,''
  arXiv:1310.3012 [hep-ph].
  %%CITATION = ARXIV:1310.3012;%%
  
\bibitem{ref:CME}  
  K.~Fukushima, D.~E.~Kharzeev and H.~J.~Warringa,
  %``The Chiral Magnetic Effect,''
  Phys.\ Rev.\ D {\bf 78}, 074033 (2008);
  %%CITATION = ARXIV:0808.3382;%%
 M.~A.~Metlitski and A.~R.~Zhitnitsky,
  %``Anomalous axion interactions and topological currents in dense matter,''
  Phys.\ Rev.\ D {\bf 72}, 045011 (2005).

\bibitem{ref:CME:cond:matt}
  A.~Vilenkin,
  %``Equilibrium Parity Violating Current In A Magnetic Field,''
  Phys.\ Rev.\ D {\bf 22}, 3080 (1980).
  %%CITATION = PHRVA,D22,3080;%%
  A.~Yu.~Alekseev, V.~V.~Cheianov, and J.~Fr\"ohlich,
 Phys. Rev. Lett. {\bf 81}, 3503 (1998).
  
\bibitem{ref:transport:related}
 P.~V.~Buividovich, M.~N.~Chernodub, D.~E.~Kharzeev, T.~Kalaydzhyan, E.~V.~Luschevskaya and M.~I.~Polikarpov,
   %``Magnetic-Field-Induced insulator-conductor transition in SU(2) quenched lattice gauge theory,''
  Phys.\ Rev.\ Lett.\  {\bf 105}, 132001 (2010);
  %%CITATION = ARXIV:1003.2180;%%
  N.~O.~Agasian,
  %``Low-energy theorems of QCD and bulk viscosity at finite temperature and baryon density in a magnetic field,''
  JETP Lett.\  {\bf 95}, 171 (2012);
  %%CITATION = ARXIV:1109.5849;%%
  S.~-i.~Nam and C.~-W.~Kao,
  %``Shear viscosity of quark matter at finite temperature in magnetic fields,''
  Phys.\ Rev.\ D {\bf 87}, 114003 (2013).
  %%CITATION = ARXIV:1304.0287;%%  

\bibitem{ref:field:estimation}
  W.~-T.~Deng and X.~-G.~Huang,
  %``Event-by-event generation of electromagnetic fields in heavy-ion collisions,''
  Phys.\ Rev.\ C {\bf 85}, 044907 (2012);
  %%CITATION = ARXIV:1201.5108;%%
  A.~Bzdak and V.~Skokov,
  %``Event-by-event fluctuations of magnetic and electric fields in heavy ion collisions,''
  Phys.\ Lett.\ B {\bf 710}, 171 (2012).
  %%CITATION = ARXIV:1111.1949;%%

\bibitem{Chernodub:2010qx} 
  M.~N.~Chernodub,
  %``Superconductivity of QCD vacuum in strong magnetic field,''
  Phys.\ Rev.\ D {\bf 82}, 085011 (2010).
  %%CITATION = ARXIV:1008.1055;%%
  %77 citations counted in INSPIRE as of 18 Aug 2013

\bibitem{Chernodub:2011mc} 
  M.~N.~Chernodub,
  %``Spontaneous electromagnetic superconductivity of vacuum in strong magnetic field: evidence from the Nambu--Jona-Lasinio model,''
  Phys.\ Rev.\ Lett.\  {\bf 106}, 142003 (2011).
  %%CITATION = ARXIV:1101.0117;%%
  %61 citations counted in INSPIRE as of 18 Aug 2013


\bibitem{Frasca:2013kka} 
  M.~Frasca,
  %``$\rho$ condensation and physical parameters,''
  JHEP {\bf 1311}, 099 (2013).
  %%CITATION = ARXIV:1309.3966;%%

\bibitem{ref:comment}
  M.~N.~Chernodub,
  %``Comment on "Charged vector mesons in a strong magnetic field",''
  Phys.\ Rev.\ D {\bf 89}, 018501 (2014).
  %%CITATION = ARXIV:1309.4071;%%
    
\bibitem{ref:Zlatko}
Z. Tesanovic, M. Rasolt and L. Xing, 
%``Quantum Limit of a Flux Lattice: Superconductivity and Magnetic Field in a New Relationship'', 
Phys. Rev. Lett. 63, 2425 (1989);
  M.~Rasolt and Z.~Tesanovic,
  %``Theoretical aspects of superconductivity in very high magnetic fields,''
  Rev.\ Mod.\ Phys.\  {\bf 64}, 709 (1992).
  %%CITATION = RMPHA,64,709;%%
  
\bibitem{ref:ferromagnetic}
  P.~Olesen,
  %``Anti-screening ferromagnetic superconductivity,''
  arXiv:1311.4519 [hep-th].
  %%CITATION = ARXIV:1311.4519;%%

\bibitem{ref:YM}
  J.~Ambjorn and P.~Olesen,
  %``On the Formation of a Random Color Magnetic Quantum Liquid in QCD,''
  Nucl.\ Phys.\ B {\bf 170}, 60 (1980);
  %%CITATION = NUPHA,B170,60;%%
%  J.~Ambjorn and P.~Olesen,
  %``A Color Magnetic Vortex Condensate in QCD,''
  Nucl.\ Phys.\ B {\bf 170}, 265 (1980).
  %%CITATION = NUPHA,B170,265;%%

\bibitem{ref:EW}
  J.~Ambjorn and P.~Olesen,
  %``On Electroweak Magnetism,''
  Nucl.\ Phys.\ B {\bf 315}, 606 (1989);
  %%CITATION = NUPHA,B315,606;%%
  J.~Ambjorn and P.~Olesen,
  %``A Magnetic Condensate Solution of the Classical Electroweak Theory,''
  Phys.\ Lett.\ B {\bf 218}, 67 (1989).
  %%CITATION = PHLTA,B218,67;%%

\bibitem{ref:earlier}
S.~Schramm, B.~Muller and A.~J.~Schramm,
  %``Quark - anti-quark condensates in strong magnetic fields,''
  Mod.\ Phys.\ Lett.\ A {\bf 7}, 973 (1992).
  %%CITATION = MPLAE,A7,973;%% 

\bibitem{ref:related:vector}
  R.~-G.~Cai, S.~He, L.~Li and L.~-F.~Li,
  %``A Holographic Study on Vector Condensate Induced by a Magnetic Field,''
  JHEP {\bf 1312}, 036 (2013);
  %%CITATION = ARXIV:1309.2098;%%
 R.~-G.~Cai, L.~Li, L.~-F.~Li and Y.~Wu,
  %``Vector Condensate and AdS Soliton Instability Induced by a Magnetic Field,''
  JHEP 1401, 045 (2014);
  %%CITATION = ARXIV:1311.7578;%%
  R.~-G.~Cai, L.~Li and L.~-F.~Li,
  %``A Holographic P-wave Superconductor Model,''
  JHEP {\bf 1401}, 032 (2014).
  %%CITATION = ARXIV:1309.4877;%%

\bibitem{ref:holographic}
  M.~Ammon, J.~Erdmenger, P.~Kerner and M.~Strydom,
  %``Black Hole Instability Induced by a Magnetic Field,''
  Phys.\ Lett.\ B {\bf 706}, 94 (2011);
  %%CITATION = ARXIV:1106.4551;%%
  N.~Callebaut, D.~Dudal and H.~Verschelde,
  %``Holographic rho mesons in an external magnetic field,''
  JHEP {\bf 1303}, 033 (2013);
  %%CITATION = ARXIV:1105.2217;%%
{
  N.~Callebaut and D.~Dudal,
  %``A magnetic instability of the non-Abelian Sakai-Sugimoto model,''
  JHEP {\bf 1401}, 055 (2014).
%  [arXiv:1309.5042 [hep-th]].
  %%CITATION = ARXIV:1309.5042;%%
}

\bibitem{ref:holographic:hexagonal}
  Y.~-Y.~Bu, J.~Erdmenger, J.~P.~Shock and M.~Strydom,
  %``Magnetic field induced lattice ground states from holography,''
  JHEP {\bf 1303}, 165 (2013);
  %%CITATION = ARXIV:1210.6669;%%
  K.~Wong,
  %``A Non-Abelian Vortex Lattice in Strongly Coupled Systems,''
  JHEP {\bf 1310}, 148 (2013).
  %%CITATION = ARXIV:1307.7839;%%
    
\bibitem{Braguta:2011hq}
  V.~V.~Braguta, P.~V.~Buividovich, M.~N.~Chernodub, A.~Y.~Kotov and M.~I.~Polikarpov,
  %``Electromagnetic superconductivity of vacuum induced by strong magnetic field: numerical evidence in lattice gauge theory,''
  Phys.\ Lett.\ B {\bf 718}, 667 (2012).
  %%CITATION = ARXIV:1104.3767;%%
  
\bibitem{Smolyaninov:2011wc}
  I.~I.~Smolyaninov,
  %``Vacuum as a hyperbolic metamaterial,''
  Phys.\ Rev.\ Lett.\  {\bf 107}, 253903 (2011).
  %%CITATION = ARXIV:1108.2203;%%

\bibitem{ref:discussion}
  M.~N.~Chernodub,
  %``Vafa-Witten theorem, vector meson condensates and magnetic-field-induced electromagnetic superconductivity of vacuum,''
  Phys.\ Rev.\ D {\bf 86}, 107703 (2012);
  %%CITATION = ARXIV:1209.3587;%%
  Y.~Hidaka and A.~Yamamoto,
  %``Charged vector mesons in a strong magnetic field,''
  Phys.\ Rev.\ D {\bf 87}, 094502 (2013);
  %%CITATION = ARXIV:1209.0007;%%
  C.~Li and Q.~Wang,
  %``Amending the Vafa-Witten Theorem,''
  Phys.\ Lett.\ B {\bf 721}, 141 (2013).
  %%CITATION = ARXIV:1301.7009;%%

\bibitem{ref:type-II:Review}
B.~Rosenstein and D.~Li, 
%{\it ``Ginzburg--Landau theory of type II superconductors in magnetic field,''}
Rev. Mod. Phys. {\bf 82}, 109 (2010).

\bibitem{Braguta:2013uc} 
{
  V.~V.~Braguta, P.~V.~Buividovich, M.~Chernodub, M.~I.~Polikarpov and A.~Y.~Kotov,
  %``Vortex liquid in magnetic-field-induced superconducting vacuum of quenched lattice QCD,''
  PoS ConfinementX {\bf 083}, (2012)
  [arXiv:1301.6590 [hep-lat]].
  %%CITATION = ARXIV:1301.6590;%%
}

\bibitem{Shushpanov:1997sf} 
{
  I.~A.~Shushpanov and A.~V.~Smilga,
  %``Quark condensate in a magnetic field,''
  Phys.\ Lett.\ B {\bf 402}, 351 (1997).
  %%CITATION = HEP-PH/9703201;%%
}

\bibitem{Bali:2012zg} 
{
  G.~S.~Bali, F.~Bruckmann, G.~Endrodi, Z.~Fodor, S.~D.~Katz and A.~Schafer,
  %``QCD quark condensate in external magnetic fields,''
  Phys.\ Rev.\ D {\bf 86}, 071502 (2012);
  %%CITATION = ARXIV:1206.4205;%%
  P.~V.~Buividovich, M.~N.~Chernodub, E.~V.~Luschevskaya and M.~I.~Polikarpov,
  %``Numerical study of chiral symmetry breaking in non-Abelian gauge theory with background magnetic field,''
  Phys.\ Lett.\ B {\bf 682}, 484 (2010).
  %%CITATION = ARXIV:0812.1740;%%
}

\bibitem{Frasca:2011zn} 
{
  M.~Frasca and M.~Ruggieri,
  %``Magnetic Susceptibility of the Quark Condensate and Polarization from Chiral Models,''
  Phys.\ Rev.\ D {\bf 83}, 094024 (2011).
  %%CITATION = ARXIV:1103.1194;%%
}

  \bibitem{Buividovich:2009ih} 
{
  P.~V.~Buividovich, M.~N.~Chernodub, E.~V.~Luschevskaya and M.~I.~Polikarpov,
  %``Chiral magnetization of non-Abelian vacuum: A Lattice study,''
  Nucl.\ Phys.\ B {\bf 826}, 313 (2010).
  %%CITATION = ARXIV:0906.0488;%%
}

\bibitem{Djukanovic:2005ag}
  D.~Djukanovic, M.~R.~Schindler, J.~Gegelia, S.~Scherer,
%{\it ``Quantum electrodynamics for vector mesons,''}
  Phys.\ Rev.\ Lett.\  {\bf 95}, 012001 (2005).
  %%CITATION = HEP-PH/0505180;%%

\bibitem{Sakurai:1960ju}
  J.~J.~Sakurai,
{\it ``Theory of strong interactions,''}
  Annals Phys.\  {\bf 11}, 1-48 (1960).

\bibitem{Chernodub:2011gs} 
  M.~N.~Chernodub, J.~Van Doorsselaere and H.~Verschelde,
  %``Electromagnetically superconducting phase of vacuum in strong magnetic field: structure of superconductor and superfluid vortex lattices in the ground state,''
  Phys.\ Rev.\ D {\bf 85}, 045002 (2012).
  %%CITATION = ARXIV:1111.4401;%%

\bibitem{Bu:2012mq} 
  Y.~-Y.~Bu, J.~Erdmenger, J.~P.~Shock and M.~Strydom,
  %``Magnetic field induced lattice ground states from holography,''
  JHEP {\bf 1303}, 165 (2013);
  %%CITATION = ARXIV:1210.6669;%%
  %19 citations counted in INSPIRE as of 09 Aug 2013
  K.~Wong,
  %``A non-abelian vortex lattice in strongly-coupled systems,''
   JHEP {\bf 1310}, 148 (2013) [arXiv:1307.7839 [hep-th]].
  %%CITATION = ARXIV:1307.7839;%%

\bibitem{Schafer:2001bq} 
  T.~Sch\"afer, D.~T.~Son, M.~A.~Stephanov, D.~Toublan and J.~J.~M.~Verbaarschot,
  %``Kaon condensation and Goldstone's theorem,''
  Phys.\ Lett.\ B {\bf 522}, 67 (2001).
  %%CITATION = HEP-PH/0108210;%%
  %90 citations counted in INSPIRE as of 21 Oct 2013

\bibitem{Miransky:2001tw} 
  V.~A.~Miransky and I.~A.~Shovkovy,
  %``Spontaneous symmetry breaking with abnormal number of Nambu-Goldstone bosons and kaon condensate,''
  Phys.\ Rev.\ Lett.\  {\bf 88}, 111601 (2002).
  %%CITATION = HEP-PH/0108178;%%
  %85 citations counted in INSPIRE as of 21 Oct 2013

\bibitem{Amado:2013xya} 
  I.~Amado, D.~Arean, A.~Jimenez-Alba, K.~Landsteiner, L.~Melgar and I.~S.~Landea,
  %``Holographic Type II Goldstone bosons,''
  JHEP {\bf 1307}, 108 (2013).
  %%CITATION = ARXIV:1302.5641;%%
  %7 citations counted in INSPIRE as of 21 Oct 2013

\bibitem{Filev:2009xp} 
  V.~G.~Filev, C.~V.~Johnson and J.~P.~Shock,
  %``Universal Holographic Chiral Dynamics in an External Magnetic Field,''
  JHEP {\bf 0908}, 013 (2009).
  %%CITATION = ARXIV:0903.5345;%%
  
\bibitem{ref:Moore}
M. A. Moore, 
%Destruction by fluctuations of superconducting long-range order in the Abrikosov flux lattice
Phys. Rev. B {\bf 39}, 136 (1989)
    
\bibitem{ref:Fischer}
M. P. A. Fisher 
%Vortex-glass superconductivity: A possible new phase in bulk high-Tc oxides
Phys. Rev. Lett. {\bf 62}, 1415 (1989).

\end{thebibliography}
\end{document}